\documentclass[12pt]{article}
\usepackage{a4wide}
\usepackage{epsfig}
\renewcommand{\baselinestretch}{1.6}
\newcommand{\msbar}{\overline{\rm MS}}
\newcommand{\msbars}{\overline{\scriptstyle\rm MS}}

\newcommand{\Li}{{\rm Li}}
\newcommand{\pfrac}[2]{\left(\frac{#1}{#2}\right)}
\newcommand{\slq}{q\kern-5.5pt/}
\newcommand{\slv}{v\kern-5pt\raise1pt\hbox{$\scriptstyle/$}\kern1pt}
\newcommand{\replabel}[1]{\vbox to0pt{\vss\hbox to0pt{\raise 24pt
  \hbox to\hsize{\hfill\rm #1}\hss}}}
\begin{document}
\begin{flushright}
MZ-TH/99-49\\
CLNS/99-1647\\
hep-ph/9911393\\
November 1999
\end{flushright}
\begin{center}
{\Large\bf $O(\alpha_s)$ corrections to the correlator\\[7pt]
  of finite mass baryon currents}\\[1truecm]
{\large S.~Groote,$^{1,2}$ J.G.~K\"orner$^1$ and
  A.A.~Pivovarov$^{1,3}$}\\[.7cm]
$^1$ Institut f\"ur Physik der Johannes-Gutenberg-Universit\"at,\\
  Staudinger Weg 7, D-55099 Mainz, Germany\\[.5truecm]
$^2$ Floyd R.~Newman Laboratory, Cornell University, Ithaca, NY 14853,
USA\\[.5truecm]
$^3$ Institute for Nuclear Research of the\\
  Russian Academy of Sciences, Moscow 117312
\vspace{1truecm}
\end{center}

\begin{abstract}
We present analytical next-to-leading order results for the correlator of 
baryonic currents at the three loop level with one finite mass quark. We
obtain the massless and the HQET limit of the correlator from the general
formula as particular cases. We also give explicit expressions for the
moments of the spectral density.
\end{abstract}

\newpage

Baryons are an important source of information on QCD interactions. Massive
baryons containing a heavy quark form a rich family of particles with many
properties experimentally measured~\cite{PDG}. The prominent object of a QCD
analysis of the properties of baryons is the correlator of two baryonic
currents and the spectral density associated with it. Calculations beyond the
leading order have not been done for many interesting cases. The massless case
has been known since long ago~\cite{pivbar}. The mesonic analogue of our
baryonic calculation was completed some time ago~\cite{Generalis} and has
subsequently provided a rich source of inspiration for many applications in
meson physics.

We report on the results of calculating the $\alpha_s$ corrections to the
correlator of two baryonic currents with one finite mass quark and two
massless quarks. We give analytical results and discuss the order of magnitude
of the $\alpha_s$ corrections. The massless and HQET limits are obtained as
special cases. We finally present explicit results on the moments of the
spectral density associated with the correlator. The discussion of the impact
of our new correlator result on baryon phenomenology will be left to future
publications.

In the present paper we limit ourselves to the simplest case with a
three-quark current of the form
\begin{equation}\label{cur}
j=\epsilon^{abc}\Psi_a(u_b^TCd_c)
\end{equation}
which has the quantum numbers of a $J^P=1/2^-$ baryon. $\Psi$ is a finite
mass quark field with mass parameter $m$, $u$ and $d$ are massless quark
fields, $C$ is the charge conjugation matrix, $\epsilon^{abc}$ is the
totally antisymmetric tensor and $a,b,c$ are color indices for the $SU(3)$
color group. Other baryonic currents with any given specified quantum numbers
are obtained from the current in Eq.~(\ref{cur}) by inserting the appropriate 
Dirac matrices. Such additions introduce no principal complication into our
method of calculation.

\newpage

\begin{figure}[htb]\begin{center}
\epsfig{figure=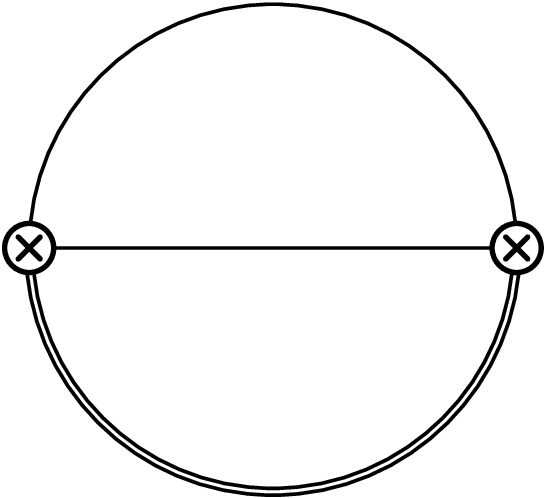, scale=0.3}\quad
\epsfig{figure=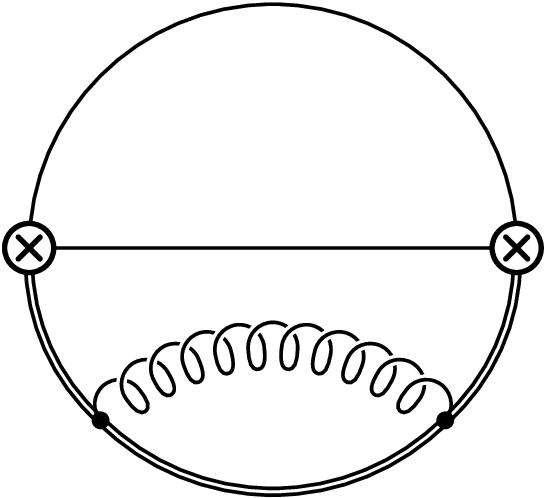, scale=0.3}\quad
\epsfig{figure=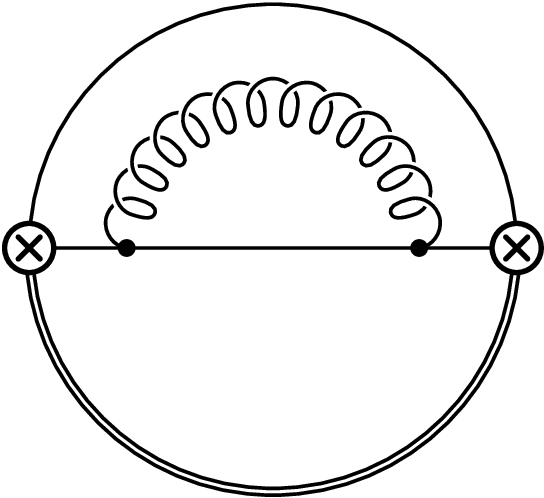, scale=0.3}\quad
\epsfig{figure=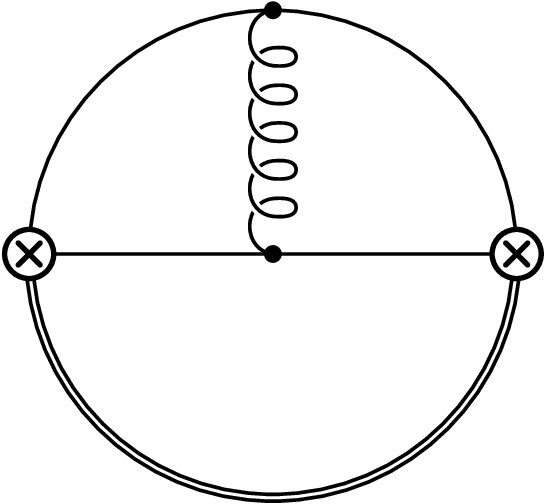, scale=0.3}\quad
\epsfig{figure=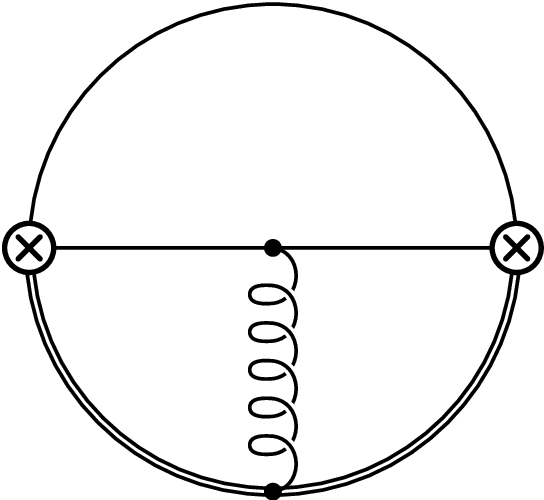, scale=0.3}\\
(a)\kern76pt(b)\kern76pt(c)\kern76pt(d)\kern76pt(e)
\caption{\label{fig1}The calculated (a) two-loop and
  (b--e) three-loop topologies}\end{center}
\end{figure}

The correlator of two baryonic currents is expanded as
\begin{equation}\label{def00}
i\int\langle Tj(x)\bar j(0)\rangle e^{iqx}dx
  =\gamma_\nu q^\nu\Pi_q(q^2)+m\Pi_m(q^2).
\end{equation}
For reasons of brevity we shall present results only for the invariant function
$\Pi_m(q^2)$. The invariant function $\Pi_m(q^2)$ can be represented compactly
via the dispersion relation 
\begin{equation} 
\Pi_m(q^2)=\int_{m^2}^\infty\frac{\rho_m(s)ds}{s-q^2}
\end{equation}
where $\rho_m(s)$ is the spectral density. All quantities are understood to be
appropriately regularized. Since the spectral density is the real object of
interest for phenomenological applications, we limit our subsequent discussion
to the spectral density instead of the correlator. To simplify formulas we
remove a trivial but awkward factor containing powers of $\pi$ and write
\begin{equation}\label{resformm}
\rho_m(s)=\frac{1}{128\pi^4}\rho(s)
\end{equation}
with a conveniently normalized $\rho(s)$,
\begin{equation}
\rho(s)=s^2\left\{\rho_0(s)\left(1+\frac{\alpha_s}\pi\ln\pfrac{\mu^2}{m^2}
  \right)+\frac{\alpha_s}{\pi}\rho_1(s)\right\}.
\end{equation}
Here $\mu$ is the renormalization scale parameter, $m$ is a pole mass of the
heavy quark (see e.g.~\cite{polemass}) and $\alpha_s=\alpha_s(\mu)$.
The leading order two-loop contribution shown in Fig.~1(a) reads
\begin{equation}\label{lead0}
\rho_0(s)=1+9z-9z^2-z^3+6z(1+z)\ln z 
\end{equation}
with $z=m^2/s$. In the $\msbar$-scheme the next-to-leading order three-loop
contribution is given by 
\begin{eqnarray}\label{corr1}
\lefteqn{\rho_1(s)\ =\ 9+\frac{665}9z-\frac{665}9z^2-9z^3
  -\left(\frac{58}9+42z-42z^2-\frac{58}9z^3\right)\ln(1-z)}\nonumber\\&&
  +\left(2+\frac{154}3z-\frac{22}3z^2-\frac{58}9z^3\right)\ln z
  +4\left(\frac13+3z-3z^2-\frac13z^3\right)\ln(1-z)\ln z\nonumber\\&&
  +12z\left(2+3z+\frac19z^2\right)\left(\frac12\ln^2z-\zeta(2)\right)
  +4\left(\frac23+12z+3z^2-\frac13z^3\right)\Li_2(z)\nonumber\\&&
  +24z(1+z)\left(\Li_3(z)-\zeta(3)-\frac13\Li_2(z)\ln z\right)
\end{eqnarray}
where $\Li_n(z)$ are polylogarithms and $\zeta(n)$ is Riemann's zeta function.
The contributing three-loop diagrams are shown in Figs.~1(b) to (e). They have
been evaluated using advanced algebraic methods for multi-loop calculations
along the lines decribed in Refs.~\cite{GKP1,GKP2}. For the convenience of
the reader we briefly sketch our method of integration. All diagrams have
first been reduced to scalar prototypes. The integrals over massless loops
have been performed (for recurrent integration where possible) and one is
left with the basic integral
\begin{equation}
V(\alpha,\beta;q^2/m^2)=\int\frac{d^Dk}{(k^2+m^2)^\alpha(q-k)^{2\beta}}
\end{equation}
which is a generalization of the standard object $G(\alpha,\beta)$ of the
massless calculation. The integral $V$ is known analytically and suffices to
calculate the diagrams in Fig.~1(c) and~(d). For the calculation of the
diagram shown in Fig.~1(e) the basic integral $V$ enters as a subdiagram.
This subdiagram then is represented in terms of a dispersion integral which
makes the whole diagram computable in terms of the same $V$ with the
argument depending on the loop momentum. The final step is a finite range
(convolution type) integration over this internal momentum with a spectral
density of the basic integral $V$. The reduction to scalar prototypes of the
diagram shown in Fig.~1(e) leads also to a new irreducible block (i.e.\ a
prototype not expressible in terms of $V$) which is related to a two-loop
master (fish) diagram. The result for this diagram is taken from
Ref.~\cite{Broad}.

The result presented in Eq.~(\ref{corr1}) is the main result of this paper:
it represents the full next-to-leading order solution. Since the anomalous
dimension of the current in Eq.~(\ref{cur}) is known to two loop
order~\cite{pivan}, the result shown in Eq.~(\ref{corr1}) completes the
ingredients necessary for an analysis of the scalar part of the correlator
in Eq.~(\ref{def00}) at the next-to-leading order level. 

We now turn to the analysis of Eq.~(\ref{corr1}). Two limiting cases of
interest are the near-threshold and high energy asymptotics. With our result
given in Eq.~(\ref{corr1}) both limits can be taken explicitly. The asymptotic
expressions can be also obtained in the framework of effective theories which
can be viewed as special devices for such calculations.

In the high energy (small mass) limit $z\rightarrow 0$ the correction reads
\begin{equation}\label{corr12}
\rho_1(s)=9+83z-4\pi^2z+2\ln z+50z\ln z+12z\ln^2z-24z\zeta(3)+O(z^2).
\end{equation}
This will lead to the small mass expansion of the spectral density in the form
\begin{equation}\label{m00}
m\rho(s)=m_{\msbars}(\mu)\rho^{\rm massless}(s)
  =m_{\msbars}(\mu)s^2\left\{1+\frac{\alpha_s}\pi
  \left(2\ln\pfrac{\mu^2}s+\frac{31}{3}\right)\right\}
\end{equation}
where $\rho^{\rm massless}(s)$ is the result of calculating the correlator in
the effective theory of massless quarks. The relation between the pole (or
invariant) mass parameter $m$ and the $\msbar$ mass $m_{\msbars}(\mu)$ reads
\begin{equation}
m=m_{\msbars}(\mu)\left\{1+\frac{\alpha_s}\pi
  \left(\ln\pfrac{\mu^2}{m^2}+\frac43\right)\right\}.
\end{equation}
Note that the massless effective theory cannot reproduce the mass
singularities (terms like $z\ln(z)$ in Eq.~(\ref{corr12})). The presence of
these singularities is an infrared phenomenon and can be parametrized with
condensates of local operators. In our case the first $m^2$ correction in
Eq.~(\ref{corr12}) (and, of course, Eq.~(\ref{lead0})) can be found if the
perturbative value of the heavy quark condensate $\langle\bar\Psi\Psi\rangle$
taken from the full theory is added~\cite{polit}. The composite operator
$\langle\bar\Psi\Psi\rangle$ should be understood within a mass independent 
renormalization scheme such as the $\msbar$-scheme. This value (perturbatively,
$\langle\bar\Psi\Psi\rangle\sim m^3\ln(\mu^2/m^2)$) cannot be computed within
the effective theory of massless quarks. It represents the proper matching
between the perturbative expressions for the correlators of the full (massive)
and effective (massless) theories. This matching procedure allows one to
restore higher order terms of the mass expansion in the full theory from the
effective massless theory with the mass term treated as a perturbation
\cite{CheSpi88}. This
is justified at high energies. Note that the correction to the next order
(i.e.\ of order $m^2/s$) can actually be found in this manner because it
depends only on one local operator and, therefore, the calculation is
technically feasible.

In the near-threshold limit $E\rightarrow 0$ with $s=(m+E)^2$ one explicitly
obtains
\begin{equation}\label{hqet0}
\rho^{\rm thr}(m,E)=\frac{16E^5}{5m}
  \bigg\{1+\frac{\alpha_s}\pi\ln\pfrac{\mu^2}{m^2}
  +\frac{\alpha_s}\pi\left(\frac{54}5+\frac{4\pi^2}9+4\ln\pfrac{m}{2E}\right)
  \bigg\}+O(E^6).
\end{equation}
The invariant function $\rho_m$ suffices to determine the complete leading
HQET behaviour since one has $\slq\rho_q+m\rho_m\rightarrow m(\slv+1)\rho$ for
the leading term. In this region the appropriate device to compute the limit
of the correlator is HQET (see e.g.~\cite{Geor,Neub}). Writing
\begin{equation}\label{hqet}
m\rho^{\rm thr}(m,E)=C(m/\mu,\alpha_s)^2\rho^{\rm HQET}(E,\mu)
\end{equation}
we obtain the known result for $\rho^{\rm HQET}(E,\mu)$~\cite{groote} with
matching coefficient $C(m/\mu,\alpha_s)$ \cite{barmatch}. In this case the
matching procedure allows one to restore the near-threshold limit of the full
correlator starting from the simpler effective theory near
threshold~\cite{Ein}.

Note that the higher order corrections in $E/m$ to Eq.~(\ref{hqet0}) can be
easily obtained from the explicit result given in Eq.~(\ref{corr1}). Indeed,
the next-to-leading order correction in low energy expansion reads
\begin{equation}\label{thre1}
\Delta\rho^{\rm thr}(m,E)=-\frac{88E^6}{5m^2}\bigg\{1+\frac{\alpha_s}\pi
  \bigg(\ln\pfrac{\mu^2}{m^2}
  +\frac{376}{33}+\frac{4\pi^2}9+\frac{140}{33}\ln\pfrac{m}{2E}\bigg)\bigg\}.
\end{equation}
It is a much more difficult task to obtain this result starting from HQET. 

We now discuss some quantitative features of the correction given in
Eq.~(\ref{corr1}). Of interest is whether the two limiting expressions
(the massless limit expression as given in Eq.~(\ref{m00}) and the HQET limit
expression in Eqs.~(\ref{hqet0}) and~(\ref{hqet})) can be used to characterise 
the full function for all energies. 

\begin{figure}[t]\begin{center}
{\epsfig{figure=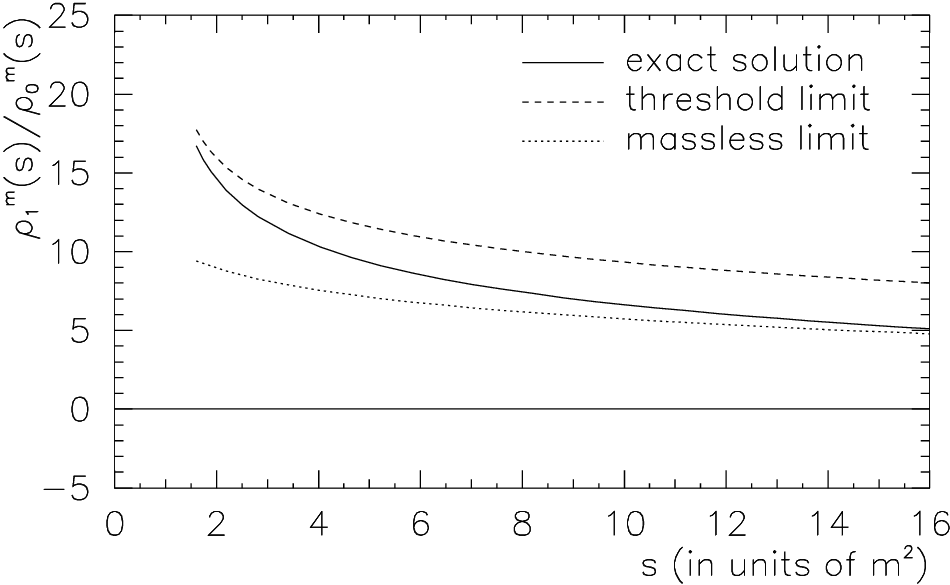, scale=0.8}}
\caption{\label{fig2}The ratio $\rho_1(s)/\rho_0(s)$ of the next-to-leading
correction and the leading order term in dependence of the energy square $s$}
\end{center}\end{figure}

In Fig.~\ref{fig2} we compare components of the baryonic spectral function in
leading and next-to-leading order. Shown is the ratio $\rho_1(s)/\rho_0(s)$.
In the following we shall always put $\mu=m$ if it is not explicitly written.
One can see that a simple interpolation between the two limits can give a
rather good approximation for the next-to-leading order correction in the
complete region of $s$.

Another informative set of observables are moments of the spectral density
\begin{equation}
{\cal M}_n=\int_{m^2}^\infty\frac{\rho(s)ds}{s^n}=m^{6-2n}M_n
\end{equation}
with $M_n$ dimensionless. We find 
\begin{equation}\label{resform1}
M_n=M_n^{(0)}\left\{1+\frac{\alpha_s}\pi
  \left(\ln\pfrac{\mu^2}{m^2}+\delta_n\right)\right\}
\end{equation}
where 
\begin{equation}
M_n^{(0)}=\frac{12}{n(n-1)^2(n-2)^2(n-3)}
\end{equation}
and
\begin{equation}
\delta_n=A_n+\frac{2\pi^2}9.
\end{equation}
The coefficients $A_n$ are rational numbers. The closed form expression for
the correction $\delta_n$ is long. Instead we present explicit expressions
for the first several moments and the differences between consecutive moments
in Table~\ref{tab1}.
We see that the difference between consecutive moments is reasonably small.
The absolute value of the correction itself in the $\msbar$-scheme has no
particular meaning and can be changed by modifying the subtraction scheme. In
normalizing to a reference moment at $n=N$, the others can be expressed
through
\begin{equation}\label{resform1norm}
\frac{M_n}{M_n^{(0)}}=\frac{M_N}{M_N^{(0)}}
\left\{1+\frac{\alpha_s}\pi(\delta_n-\delta_N)\right\}.
\end{equation}
One now can find the actual magnitude of the correction from Table~\ref{tab1}.
Indeed, for any given precision and range of $n$ the set of perturbatively
commensurate moments can be found.

Note that moments represent massive vacuum bubbles, i.e.\ diagrams without
external momenta with massive lines. These diagrams have been comprehensively
analyzed in Refs.~\cite{avdeev,david}. The analytical results for the first
few moments at three-loop level can be checked independently with existing
computer programs (see e.g.~\cite{ChKStmom8}).

Moments of the spectral density are convenient observables for
phenomenological applications. Exact results for correlators with massive 
particles are not known in higher orders for many important physical channels.
Therefore one generally uses approximate procedures. We consider the two
formal limiting cases as a base for possible interpolation. Both are simpler
than the full calculation.

For the first approximation, the high energy or massless approximation, we
first check the numerical difference between the exact result and the massless
approximation. The massless limit for the moments means that one formally
integrates the expression $\rho^{\rm massless}(s)$ in Eq.~(\ref{m00}) over the
range $(m^2,\infty)$ to obtain
\begin{equation}
{\cal M}_n^{\rm massless}=\int_{m^2}^\infty\frac{\rho^{\rm massless}(s)ds}{s^n}
  =m^{6-2n}M_n^{\rm massless}.
\end{equation}
The result is
\begin{equation}
M_n^{\rm massless}=\frac1{n-3}\left\{1+\frac{\alpha_s}\pi
  \left(2\ln\pfrac{\mu^2}{m^2}+\frac{31}3-\frac2{n-3}\right)\right\}.
\end{equation}
This is the extrapolation from the side of high energies. The opposite limit
for the moments is obtained by using the near-threshold spectral density for
the integration along the whole $s$-axis. This approximation is an
extrapolation from the threshold region which is clearly a poor approximation
far from threshold. However, it formally exists for sufficiently large $n$ and
can be obtained with the simple HQET result shown in Eqs.~(\ref{hqet0})
and~(\ref{hqet}). The moments are given by
\begin{equation}
M_n^{\rm thr}=m^{2n-6}\int_{m^2}^\infty\frac{\rho^{\rm thr}(E)ds}{s^n}
  =2m^{2n-6}\int_0^\infty\frac{\rho^{\rm thr}(E)dE}{(m+E)^{2n-1}}.
\end{equation}
From Eq.~(\ref{hqet0}) we have
\begin{equation}
M_n^{{\rm thr}(0)}=12\frac{2^6(2n-8)!}{(2n-2)!}
  =\frac{12}{n^6}\left(1+O(n^{-1})\right)
\end{equation}
where the large $n$ result specifies the region of applicability of the
near-threshold approximation for the moments of the spectral density. The
correction is 
\begin{equation}
M_n^{\rm thr}=M_n^{{\rm thr}(0)}\bigg\{1+\frac{\alpha_s}\pi
  \bigg(\ln\pfrac{\mu^2}{m^2}
  +\frac{54}5+\frac{4\pi^2}9-4\ln 2-4(\psi(2n-7)-\psi(6))\bigg)\bigg\}
\end{equation}
where $\psi(z)$ is Euler's $\psi$-function.

One can see that the corrections to the moments basically reflect the shape of
the correction to the spectrum as given in Eq.~(\ref{corr1}). The massless
approximation is reasonably good for relative corrections for the first few
moments  despite the unfavorable shape of the weight function $1/s^n$. It can
be improved by changing the subtraction point $\mu$ or by resumming the
integrand~\cite{pivtau} which lies beyond the scope of finite order
perturbation theory though.

To conclude, we have computed the next-to-leading perturbative corrections to
the finite mass baryon correlator at three-loop order. Technically, the method
allows one to obtain analytical results for two-point correlators of composite
operators with one finite mass particle which can be compared to HQET results.
Corrections in $E/m$ near threshold are easily available from our explicit
results. From threshold to high energies the exact spectral density
interpolates nicely between the leading order HQET result close to threshold
and the asymptotic mass zero result. Going even one order higher it is very
likely that the full four-loop spectral density can be well approximated by
the corresponding massless four-loop result which can be calculated using
existing computational algorithms~\cite{ibyparts,CheSmi}.

\subsection*{Acknowledgements}
The present work is supported in part by the Volkswagen Foundation under
contract No.~I/73611 and by the Russian Fund for Basic Research under contracts
Nos.~97-02-17065 and 99-01-00091. A.A.~Pivovarov is an Alexander von Humboldt
fellow. S.~Groote gratefully acknowledges a grant given by the Max Kade
Foundation.

\begin{table}\begin{center}
{\renewcommand{\baselinestretch}{0.625}
\begin{tabular}{rcl}&&\\
  $n$&$A_n$&$\delta_n-\delta_{n-1}$\\&&\\\hline&&\\
 4&3&\\
 5&13/2&3.500000\\
 6&17/2&2.000000\\
 7&535/54&1.407407\\
 8&1187/108&1.083333\\
 9&64093/5400&0.878333\\
10&22691/1800&0.737037\\
11&1167767/88200&0.633878\\
12&2433499/176400&0.555357\\
13&68055703/4762800&0.493665\\
14&14034047/952560&0.443968\\
15&348916495/23051952&0.403114\\
16&8935543717/576298800&0.368960\\
17&3086442535481/194788994400&0.340003\\
18&3147830736641/194788994400&0.315152\\
19&3205021446217/194788994400&0.293603\\
20&6517078055669/389577988800&0.274746\\
21&382499185005589/22517607752640&0.258112\\
\end{tabular}}
\vspace{12pt}
\caption{\label{tab1}Values for the rational part $A_n$ of the first moments
  $\delta_n$ and their relative difference $\delta_n-\delta_{n-1}$}
\end{center}\end{table}

\end{document}